\documentstyle[preprint,aps,epsf]{revtex}

\begin{document}
\title{Electron interactions, classical integrability, and
level statistics in quantum dots}
\author{Lilia Meza-Montes and Sergio E. Ulloa}
\address{Department of Physics and Astronomy,
Condensed Matter and Surface Science Program, Ohio University,
Athens, OH 45701-2979, USA}
\author{Daniela Pfannkuche}
\address{Max-Planck Institut f\"ur Festk\"orperforschung,
Heisenbergstr.\ 1, D--70569 Stuttgart, Germany}
\date{August 28, 1997}
\maketitle

 \begin{abstract}
 The role of electronic interactions in the level structure of
semiconductor quantum dots is analyzed in terms of the correspondence
to the integrability of a classical system that models these
structures.  We find that an otherwise simple system is made strongly
non-integrable in the classical regime by the introduction of particle
interactions.  In particular we present a two-particle classical system
contained in a $d$-dimensional billiard with hard walls.  Similarly, a
corresponding two-dimensional quantum dot problem with three particles
is shown to have interesting spectral properties as function of the
interaction strength and applied magnetic fields. 
 \end{abstract}

\hfill
\hfill\\

\noindent
Sergio E. Ulloa\\ 
E-mail: sulloa1@ohiou.edu\\
Fax: (614) 593--0433


\normalsize

\section{Introduction}

The role of electronic interactions on the observed properties of
semiconductor quantum dots has increasingly been found to be of vital
importance, as many of the papers in these proceedings and elsewhere
illustrate.  More and more, transport and capacitance experiments, as
well as detailed studies of far-infrared response, require the
consideration of interactions in order to understand the observed
experimental features.  Since a number of experiments explore
specifically the electronic level spectrum, it is of interest to
contrast these studies with a detailed theoretical analysis of the role
of interactions.  We present here an analysis of the energy level
statistics of a quantum dot system as function of variable interaction
strength, and then as function of an applied magnetic field.

As the level spectrum is considered in detail, the correspondence with
the dynamical integrability of a classical system is also of interest. 
It is anticipated that as the Coulomb interaction is introduced, the
dynamics would in general become chaotic and this would in turn be
reflected in various statistical properties of the level spectrum.  It
has been known for some time now that as a classical system becomes
non-integrable \cite{Gutzwiller}, the corresponding quantum system
exhibits a level spacing distribution given by the `Wigner surmise'
derived in the context of random matrix theory (RMT) \cite{Mehta}.  In
fact, this behavior has been verified in a number of theoretical and
experimental systems, although typically the classical
non-integrability is due to disorder or boundary (geometrical) effects.
 Examples discussed in the literature include small disordered metallic
particles \cite{Rusky}, particles moving in a variety of `stadia'
\cite{Heller,Stone}, and in two-dimensional antidot arrays
\cite{Weiss,Ketzmerick,Surkhe-prb}.  RMT also has been used to make
definite predictions on the statistical distribution of Coulomb
blockade peak amplitudes \cite{Jalabert-etal}. This behavior has in
fact been shown recently to be a good description of quantum dot
systems in beautiful experimental realizations \cite{Marcus-Chang}. 
One should also mention that recent work on interacting systems, some
without an obvious classical counterpart, has also shown that these
exhibit the level repulsion characteristic of quantum versions of
non-integrable classical systems \cite{Bruus-Hubb}.   Moreover, recent
analysis of the level spectrum of excitons in quantum wells (via
photoluminescence excitation spectroscopy) has shown evidence of level
repulsion \cite{Vina-ICPS}.

 A level structure described by RMT has been sought recently in the
charging (or addition) spectrum of quantum dots \cite{Sivan-Simmel}. 
Detailed analysis of experimental charging energies, after proper
subtraction, would be expected to yield the single-particle--like {\em
excitation} spectrum described by RMT.\@  Unfortunately, it appears
that the extraction of this excitation spectrum is obscured by the
systematic shift in the charging energies, and the resulting level
spacing is described by a normal distribution, rather than by the RMT
functions. This would then suggest that a more direct measurement of
the excitation spectrum (via perhaps nonlinear tunneling currents)
would be desirable (although clearly difficult experimentally beyond
the first few excitations \cite{Johnson-McEuen-Weis}).  One should also
mention here that it is believed that the nonlinear transport
experiments explore mainly the excitations of the center of mass of the
system (in the typically parabolic quantum dots) \cite{PU-PRL-FKP}, due
to the strong electronic correlations suppressing most of the tunneling
`channels'.
 This prevalence of the center of mass excitations is however expected
theoretically to diminish as the energy of the excitation increases
\cite{PU-PRL-FKP}.  This regime would be reached only as the bias
voltage is raised in transport experiments, and makes it then difficult
to achieve in practice.  We hope, however, that the results presented
here would motivate more experimental work in this direction.

\section{A classical dot}

We have studied the effects of particle interaction on the classical
integrability of a system of two masses  moving inside a
$d$-dimensional billiard.  We find, in general, that the motion is
strongly chaotic (typically exhibiting `soft' chaos, with a mixed
dynamics, where regions of phase space are still periodic or
quasi-periodic), and with a strong dependence on the characteristic
interaction length and strength.  As perhaps one of the simplest
examples (see Ref.\ \cite{lilia-RC} for a description of the general
case), consider two particles of equal masses moving in a 1D box
defined in the interval $q:(- {1\over 2}, {1\over2})$ (we measure all
lengths in terms of the box size).  The particles are assumed to
interact via a screened potential
$ V(q_1, q_2) = \exp (- \lambda |q_1 - q_2 |) / |q_1 - q_2| $ ,
 where $\lambda$ is the inverse screening length.  Notice that this
potential goes to a hard-core $\delta-$function when $\lambda
\rightarrow \infty$, and the particles behave then as non-interacting
but impenetrable points.  In that case, the dynamics can be integrable
in special cases of the mass ratios \cite{Sinai-Kozlov}.  In this
sense, $\lambda$ plays the role of a perturbation parameter which
changes the degree of integrability of the system, as it determines the
effective `radius' of the particles for a given total energy.  Direct
calculations show that this is indeed the case.

 The problem could in principle be solved by direct integration of the
equations of motion, but we find convenient to transform it to a set of
center-of-mass and relative coordinates $ R  =  (q_1 + q_2)/M  $, and
$r = q_2 - q_1$, respectively, where the total mass $M=2$, and the
reduced mass $\mu = 1/2$. These equations define a two-dimensional
space of coordinates ${\bf \rho} = (r,R)$. In this space, we have a new
set of equations for the boundary of the billiard, say $F_j(\rho)$.
The Hamilton equations are transformed to $\dot{ r}  = p/\mu  $,
$\dot{R}   = P/M $, and
 \begin{eqnarray}
  \dot{p} & = & -\nabla_{r} V(r) +  
     \sum _{j} {A}_j ({ p,P}) \, \delta[F_j({r,R})]   \nonumber \\
 \dot{ P} & = & \sum _{j} { B}_j({p,P}) \, \delta[F_j({r,R})]. 
 \end{eqnarray}
 The functions ${ A}_j$ and ${B}_j$ describe the change in the momenta
$p$ and $P$, due to the bounce on the $j$-th wall. Notice that these
equations describe the motion of a  single `hyperparticle' in  the
two-dimensional $\rho$-space.   This {\em hyperbilliard} description
can be generalized to any number of dimensions \cite{lilia-RC}.

Notice that bounces of the hyperparticle in the hyperbilliard
correspond to bounces of the masses in the real/original dot. The walls
of the billiard cause the breaking of translational symmetry of the
system, and as a consequence, the center-of-mass (CM) momentum is no
longer a constant of motion. In the case of non-interacting and
equal-mass particles, the changes in the CM momentum are determined
only by the geometry of the billiard.  In our case, however, the
interaction couples the CM and relative momenta after each bounce,
which in turn depend on the momenta of each of the original masses. 

{\bf Dynamical map}. --- We should also mention that these equations in
$\rho$-space provide an interesting description which is also extremely
useful: In between bounces, the hyper-particle moves freely along the
$R$-axis, whereas the interaction acts only along $r$.  These two
motions are independent, and only become correlated at each bounce,
when the different momentum components are changed, while keeping the
total energy constant.  Understanding this fact allows one to describe
the motion in terms of a {\em dynamical map} connecting the different
bounces.   If the coordinate of the hyperparticle is $\rho_n = (r_n,
R_n)$ at the time of the $n$-th bounce, then the time spent until the
next bounce on the $j$-th wall is obviously the same along {\em both}
components, and one can then write
 $ \tau_r (  \rho_n , \rho_{n+1}  ) = \tau_{R} ( \rho_n , \rho_{n+1} ) $.
 Since the CM motion is that of a free particle, except for the
collisions with the walls, $\tau_R$ can be calculated simply.
Similarly, for a pure Coulomb potential ($\lambda = 0$), $\tau_r$ can
be calculated analytically, so that the previous equation can be
written in a more explicit form,
 $ \tau \{ T_\epsilon ( \rho_n ), T_\epsilon ( \rho_{n+1} ) \} 
 = \mid  ( R_{n+1} - R_n ) / (P/M) \mid $,
 where $\tau$ is now the time elapsed going from $\rho _n$ to
$\rho_{n+1}$, expressed in terms of the time $T_\epsilon (\rho)$ spent
by the particle from the turning point to $\rho$,
 $ T _\epsilon (\rho) = \frac{ rp}{2 \epsilon} + \left(\frac {\mu}{2
 \epsilon ^3} \right)^{1/2} \cosh^{-1} ( r  \epsilon) ^{1/2} $.
 Here $\epsilon = E -P^2/2M$, represents the energy left for the
relative motion, as $E$ is the total energy of the two-mass system.
 For a weakly screened potential, 
$\lambda \leq 1$, we can expand $V(r)$ to first order and obtain
a similar expression, where $E$ is scaled to $E- \lambda$.  What
follows, after these definitions is to characterize all the different
possible trajectories in the triangular region in $\rho$-space. A
simple algorithm can then be obtained to determine the Poincar\'e
surfaces of section.  This nontrivial (and clearly nonlinear) algebraic 
map provides then a
full description of the dynamics.  Its use (in lieu of the direct integration
of the equations of motion) simplifies calculations a
great deal, and allows one to better characterize the system, as we
describe below.  

{\bf Poincar\'e sections}. ---
 In order to characterize the motion of the two interacting particles
(or the hyperparticle with its many degrees of freedom), we explore
Poincar\'e sections of the resulting phase space.  Having a
four-dimensional space in this case $(r,R,p,P)$, we select to show
those where one of the particles is at one end of the box (notice that
the energy is a constant of motion here).  Figure 1 shows a typical
Poincar\'e section for $\lambda = 0.1$ and $E=1.5$, obtained using the
map described above (and which is virtually identical to the one
obtained directly from integration of the equations of motion
\cite{lilia-RC}).  Notice that chaotic trajectories nearly fill the
available phase space (for this given total energy).  Increasing values
of $\lambda$ give chaotic orbits that fill more of the available phase
space.  Moreover, there are also a number of islands of stability, as
expected from the KAM theorem \cite{Gutzwiller}, near the fixed point
corresponding to periodic symmetric motion in the non-interacting
system (and indicated in the figure with a cross on the right axis). 
In fact, for all values of $E$ and $\lambda$, the initial condition
where $q_1(0) = - q_2 (0) = \frac{1}{2}$, and $p_1(0)= -p_2(0) = \sqrt
{E - V(1)}$ gives rise to a periodic orbit. Other islands also appear
purely due to the interaction, as can be seen near $q_2 = 0.2, p_2=0$,
and are associated with a sort of correlated motion of the two masses.

General values of $E$ and $\lambda$ give rise to this type of mixed
dynamics, with chaotic and regular trajectories sharing the available
phase space.  This increasing degree of `soft chaos' in the system (as
$\lambda$ increases, for example), should be reflected in the level
statistics of the corresponding quantum mechanical system, as we
describe explicitly elsewhere in these proceedings. One naturally
expects that this effect of interactions --- turning a regular system
into a non-integrable one --- would be rather pervasive, regardless of
the type of particle confinement and details of the interactions.  In
the following section we illustrate this effect in a somewhat different
quantum mechanical model of a quantum dot.

\section{A quantum dot}

Here, we will use a model of a quantum dot which has been very
successful in the description of recent experiments in these structures
\cite{Weiss,PU-PRL-FKP,Weinmann}.  The quantum dot is modeled
as a parabolic potential well with circular symmetry, and the frequency
(or curvature) is chosen to fit characteristic single-particle
excitation energies in these devices.  In this case, the few-particle
problem, which includes fully the effect of interactions, can be solved
quite accurately (numerically `exactly') even in the presence of
magnetic fields.  The approach is based on a canonical (Jacobi)
transformation to a set of auxiliary harmonic oscillator generating
operators which allow one to write the interaction matrix elements in a
closed analytical form, easily calculable \cite{PH-DP-prl93}.  This,
together with the ability to separate the center of mass (CM) degree of
freedom from the `relative' ones (a feature almost exclusive to the
parabolic confinement potential), allows one to completely characterize
and solve for the spectrum of up to three electrons in this parabolic
well. Correspondingly, the spectrum can be characterized by 
 $ E = \varepsilon (N_{CM}, M_{CM}) + E_{\rm rel} $,
 where the first term gives the CM manifold, and $E_{\rm rel}$ is
obtained from the relative-motion part of the Hamiltonian.  In the case
of three particles, this  latter part has diagonal elements given by
 $ \varepsilon (N_- , M_-) + \varepsilon (N_+ , M_+)$,
which are mixed by the Coulomb interaction (taken here as given by
$V(r_{ij}) = e^2/(\epsilon r_{ij})$, with $\epsilon$ a background
dielectric constant which would be assumed variable in the results
section below).  Notice that in all these expressions,
 \begin{equation}
 \varepsilon ( N, M) = \hbar \Omega_+ \left( N + \frac{1}{2} \right) +
 \hbar \Omega_- \left( M + \frac{1}{2} \right) \, ,
 \end{equation}
 arise from the auxiliary harmonic oscillators introduced in the
canonical transformation \cite{PH-DP-prl93}, with $\Omega_{\pm} = \left
[ \sqrt{ 4 \Omega_0^2 + \omega_c^2} \pm \omega_c \right ] /2 $. Here,
$\Omega_0$ characterizes the single-particle harmonic confinement
potential of the quantum dot (with a typical value of $\hbar \Omega_0 
\approx $ 1--2 meV), and $\omega_c = eB/mc$ is the cyclotron frequency
of the electron.  Due to the rotational symmetry of the system, the
Coulomb interaction only couples states with the same relative angular
momentum, given by $L_{\rm rel} = \hbar (M_+ + M_- - N_+ - N_-)$. This is
important, for apart from making the calculation simpler,  it also
allows us to carry out matrix diagonalizations with extremely high
accuracy and for very many eigenstates (typically one to two thousand
levels).  This complete convergence is obviously important if one is
interested in the analysis of the level spacings, as we now proceed to
do.

{\bf Level statistics results}. ---
 The statistical analysis of the levels obtained as described above for
three particles can be carried out according to the typical
prescriptions in the literature.  The level spectrum is used to extract
a slowly-varying density of states with a smooth energy dependence,
characteristic of the system at hand. This {\em unfolding} procedure
then leaves one to study the structure of the level fluctuations of the
spectrum, on which a number of statistical tests and statements can be
made \cite{Mehta,Gutzwiller}.  The unfolding here is performed by
defining the `staircase' function $N(E)$ which gives the cumulative
number of states below $E$, and then fitting this to a smooth
polynomial (typically of fourth degree), $\tilde{N}(E)$. The
`linearized' or unfolded level sequence is then obtained from $x_i =
\tilde{N} (E_i)$, where $E_i$ is the original sequence.  This process,
although not unique, gives similar results to other unfolding
procedures (see \cite{Bruus-Hubb} for a good discussion).
 
One can perform a number of statistical analyses.  Here we focus on the
nearest-neighbor spacings (NNS), obtained from $s_i = x_i - x_{i-1}$,
and calculate the probability density function $P(s)$ for a given
sequence.  It is useful to define the integrated probability function
\cite{Robnick}, $ I(s) = \int_0^s ds' P(s') $. Notice that $I(s)$ is
nothing but the total number of NNS below a given $s$, and can be
uniquely calculated, without any dependence on the specific binning
used to calculate the typical histogram representations of $P(s)$. In
what follows, we use $I(s)$ to make quantitative statements, but revert
to showing the more conventional $P(s)$.

Figure 2 shows a sequence of $N(E)$ staircase curves for different
values of the background dielectric constant $\epsilon$, defined above.
The sequences analyzed have all the same value of relative angular
momentum, $L_{\rm rel} = 3\hbar$, and total spin $S=3\hbar /2$, and for
a magnetic field value of $B= 3$T.\@  Increasing values of $\epsilon$
would produce a progressively weaker value of the Coulomb interaction,
and vice versa.  We have varied $\epsilon$ such that the interaction
varies by up to a factor of 50.  In the figure, the curve labeled 1
corresponds to the value of $\epsilon = \epsilon_1= 12.5$, found in
GaAs, where most quantum dots are defined.  Increasing the effective
Coulomb interaction (with $\epsilon = \epsilon_1/10$), produces the
smoothest staircase function (labeled 10), completely devoid of the
harmonic oscillator `steps' in weaker interactions (as seen here for
relative interaction strength 1/5 and 1/3).  One expects that for
interaction 10, the level mixing produced would be substantial,
resulting in a level structure well described by the RMT distributions,
as seen in other systems.  We find that this is not the case here.

In fact, if one analyzes the appropriate histograms for the NNS
distribution function $P(s)$, the interaction strength shows its effect
quite clearly, as shown in Fig.\ 3.  There, we show the corresponding
$P(s)$ for three different interaction strength (or $\epsilon$) values.
 The anticipated Poisson distribution function one obtains for a
`generic' integrable system \cite{Mehta,Gutzwiller} is similar to the
case for interaction 1/3.  However, notice that this $P$ here goes to
zero even {\em faster} than the expected Poisson form, reflecting the
non-generic character of the delta-function--like distribution of the
pure harmonic oscillator system. Moreover, as the
interaction becomes stronger, notice that the distribution functions
have a maximum not at zero $s$, but rather at a finite value.  This
behavior is more in agreement with the RMT predictions, where $P$ would
be expected to be given by a GUE function (given the finite magnetic
field), where $P(s \approx 0) \propto s^2$. One should notice, however,
that in all cases we have studied, $P(s)$ never fully reaches the
anticipated GUE for a fully chaotic system. The reason for this lack of
full crossover into the GUE distribution is perhaps associated with the
mixed dynamics of the corresponding classical system.  In such cases,
it has been argued that the level distribution can be seen as a
superposition of Poisson and GUE functions, with weights corresponding
to the coverage in phase space for the integrable and non-integrable
regions, respectively \cite{Robnick}.  At this point, however, we have
not analyzed the classical system that directly models this quantum
dot, and expect to report on this relation elsewhere.  However, it may
also be the case that this incomplete crossover to a GUE arises from
the peculiar non-generic non-Poissonian distribution function of the
integrable (non-interacting) system composed by overlapping harmonic
oscillators, and/or remnant hidden symmetries.  We have also fitted the
$I$-functions to the well known Brody distribution, $I_{\rm Brody} = 1 
- \beta_\alpha \exp (-s^{\alpha +1})$, providing a (phenomenological)
measure of the crossover ($\alpha =0$ for Poisson, $\alpha=1$ for GOE).
In Fig.\ 2 and for unit interaction strength, we get $\alpha = 0.04$,
while $\alpha = 0.25$ for interaction 10.

Furthermore, the non-typical character of the level distribution in the
harmonic oscillator is also reflected in the magnetic field dependence
\cite{SU-DP-StaFe}.  If one studies $P(s)$ for a given interaction
strength but increasing magnetic fields, the system has a non-monotonic
evolution.  Examples of this behavior are shown in Ref.\
\cite{SU-DP-StaFe}.  For low magnetic field ($B =0.3$ T, and `unit'
interaction, with $\epsilon = \epsilon_1$), $P(s)$ is not the harmonic
oscillator delta-like function, but is somewhat shifted towards a GUE
(Brody-fit with $\alpha \approx 0.1$). As the field increases, however,
the $P$-distribution reaches a `maximum' crossover ($\alpha \approx
0.2$, for $B=1$ T), before going back towards a more delta-like
function, as seen for $B=10$ T, which drops faster than a Poissonian
distribution.  That this occurs is understandable, since for high
magnetic fields one would expect to reach a regime where the
interactions would be a weak perturbation (for a given value of
$\epsilon$), and the spectrum would evolve towards a set of Landau
levels. This non-monotonic behavior is seen in all the level sequences
we have studied.

\section{Conclusions}

We have shown that the interactions introduce classical
non-integrability in a quantum-dot system, even if the geometry is
integrable for the one-particle problem.  Moreover, this behavior,
given the Bohigas {\em et al}. conjecture \cite{Gutzwiller},
would be expected to be reflected in the accompanying level structure
of the corresponding quantum mechanical version of the system.  We have
found, that as we introduce interactions, indeed the NNS fluctuation
distribution function exhibits an apparent crossover towards one of the
RMT functions, with a characteristic maximum away from zero spacing. 
This crossover, is found to be not complete, however, even for rather
strong interaction, perhaps due to the mixed dynamics in the classical
counterpart.  The NNS distribution exhibits non-monotonic magnetic
field dependence, associated with the fact that in that regime the
particle interactions are but a weak perturbation of the Landau level
spectrum.  A number of theoretical questions still remain, however, as
more direct comparison of the classical and quantum systems is needed,
perhaps including a study of the associated wave functions and possible
appearance of `scars' \cite{Heller}.  Nevertheless, the importance of
interactions in the details of level structure and associated
experiments, as discussed in the introduction, cannot be ignored.  We
hope that this, as well as semiclassical treatments (see Richter's
article in these proceedings), motivate further understanding of this
fascinating problem. 

\vspace{1ex}
{\bf Acknowledgments}.
This work was supported in part by US DOE grant no.\ DE-FG02-91ER45334,
CONACyT-M\'exico, and the Deutsche Forschungsgemeinschaft.  SEU also
acknowledges support from the A. von Humboldt Stiftung.

\begin{figure}[htb]
\caption{Poincar\'e section of two-particles in a box with interaction
given by $\lambda = 0.1$.  Total energy $E=1.5$, and $q_1=-\frac{1}{2}$.}
\end{figure}

\begin{figure}[htb]
\caption{Typical staircase function for different interaction strengths.
For interaction strength 10, the harmonic oscillator structure is 
completely lost, while it is visible for weak interaction (1/3 and 1/5).}
\end{figure}

\begin{figure}[htb]
\caption{NNS distribution function for varying interaction
strength values (in units of the interaction in GaAs).  Notice
evolution towards a GUE distribution for stronger interaction.}
\end{figure}

\begin{figure}[htb]
\epsfxsize=4.5in
\epsfbox{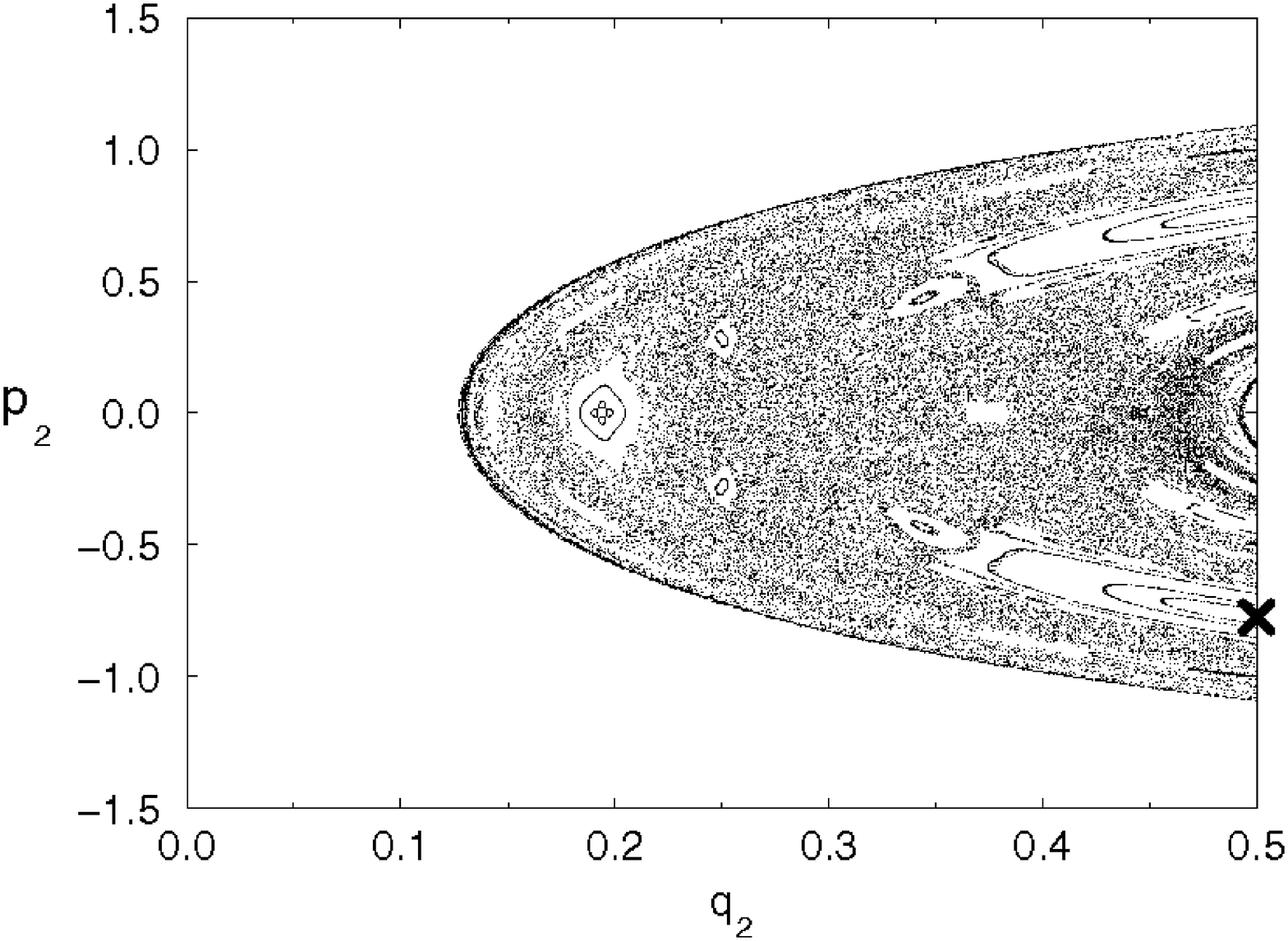}
\end{figure}
{\large FIG 1}

\begin{figure}[htb]
\epsfxsize=4.5in
\epsfbox{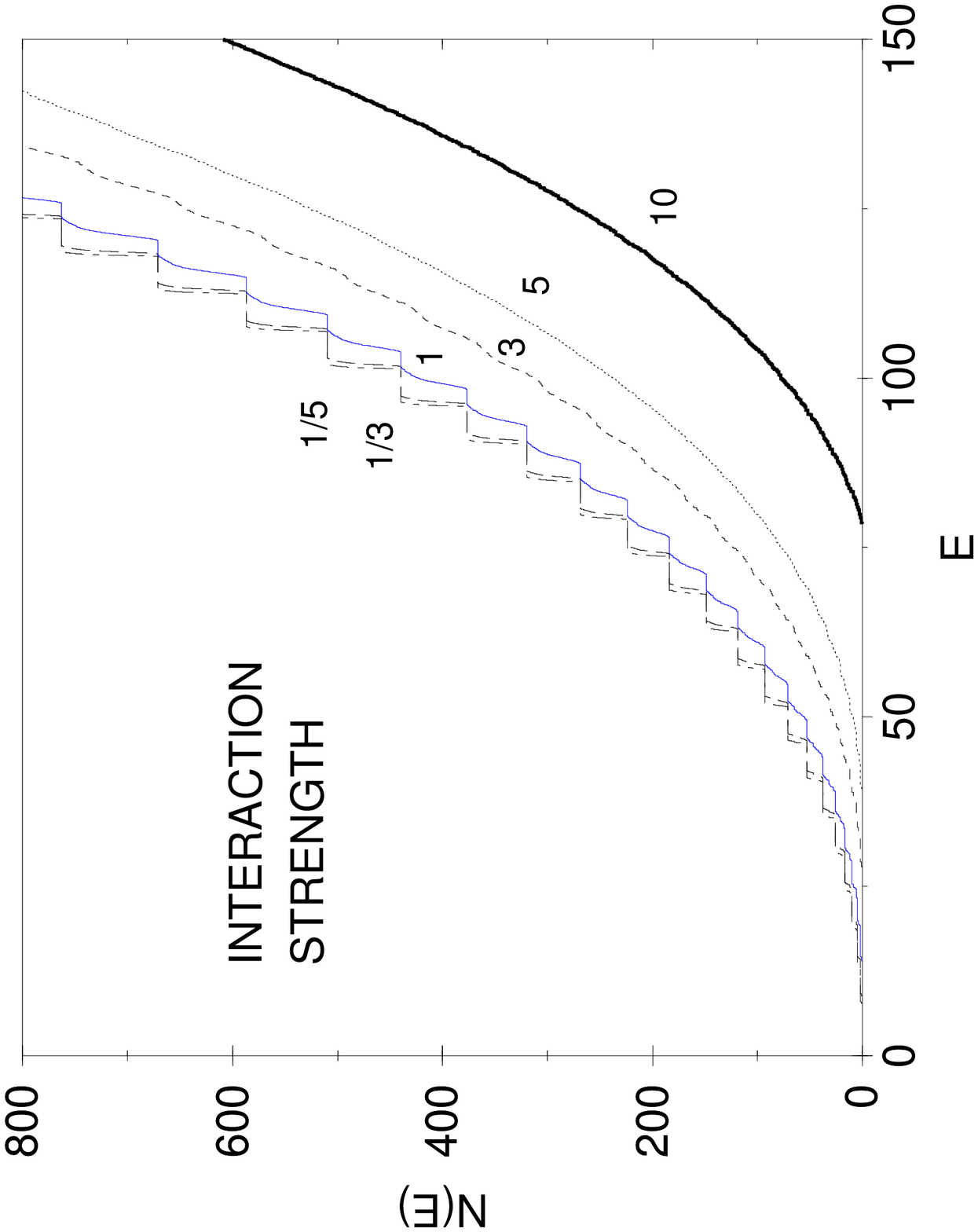}
\end{figure}
{\large FIG 2}

\begin{figure}[htb]
\epsfxsize=4.5in
\epsfbox{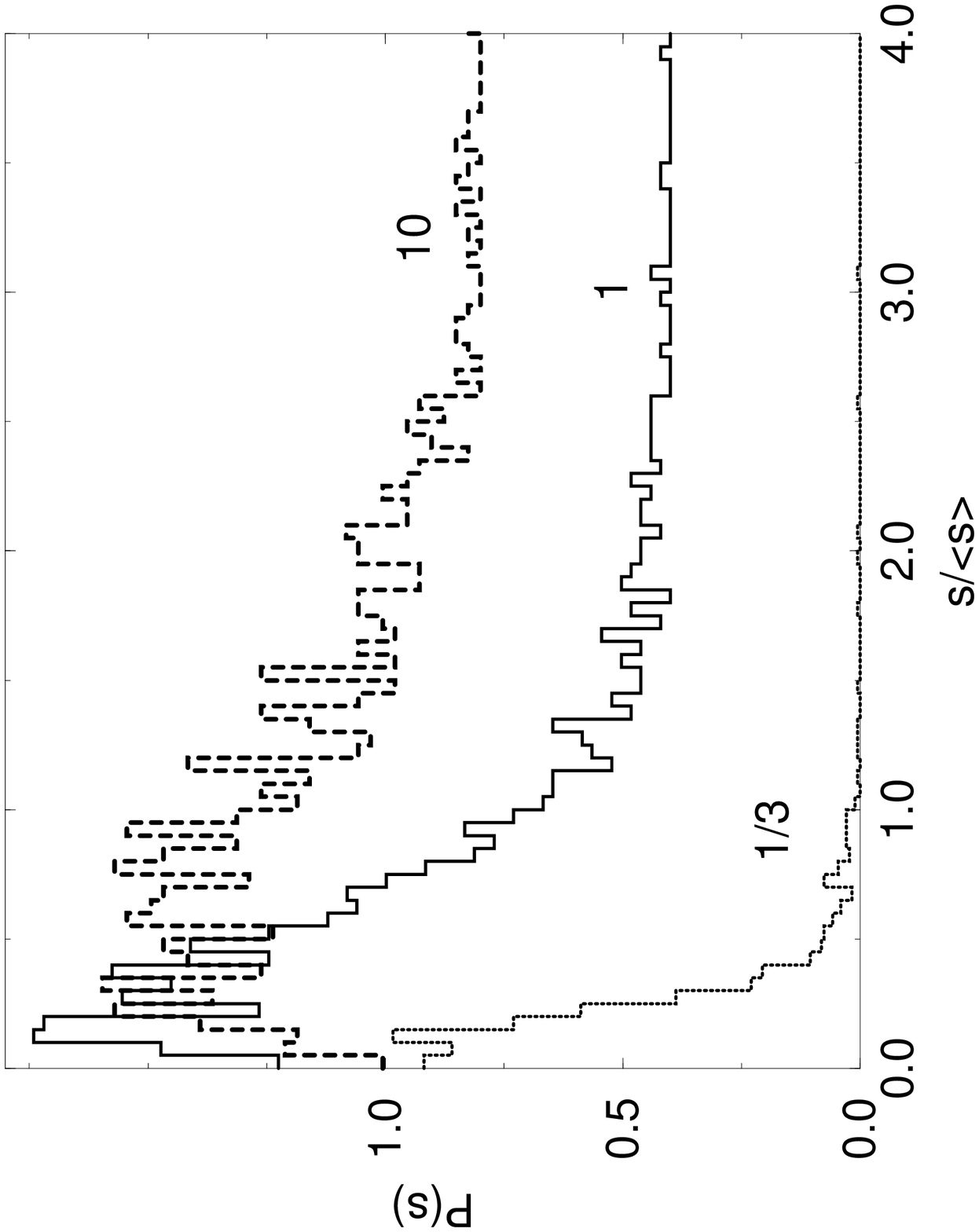}
\end{figure}
{\large FIG 3}

\end{document}